# Quantum-inspired detection for Spectral Domain Optical Coherence Tomography

SYLWIA M. KOLENDERSKA[1,2,*], FRÉDÉRIQUE VANHOLSBEECK[1,2], AND PIOTR KOLENDERSKI[3]

[1]*The Dodd-Walls Centre for Photonic and Quantum Technologies, New Zealand*
[2]*The Department of Physics, The University of Auckland, Auckland 1010, New Zealand*
[3]*Faculty of Physics, Astronomy and Informatics, Nicolaus Copernicus University, Grudziądzka 5, 87-100 Toruń, Poland*
[*]*Corresponding author: skol745@aucklanduni.ac.nz*

**The intensity levels allowed by safety standards (ANSI or ICNIRP) limit the amount of light that can be used in a clinical setting to image highly scattering or absorptive tissues with Optical Coherence Tomography (OCT). To achieve high-sensitivity imaging at low intensity levels, we adapt a detection scheme - which is used in quantum optics for providing information about spectral correlations of photons - into a standard spectral domain OCT system. This detection scheme is based on the concept of Dispersive Fourier Transformation, where a fibre introduces a wavelength-dependent time delay measured by a single-pixel detector, usually a high-speed photoreceiver. Here, we use a fast Superconducting Single-Photon Detector (SSPD) as a single-pixel detector and obtain images of a glass stack and a slice of onion at the intensity levels of the order of 10 pW. We also provide a formula for a depth-dependent sensitivity fall-off in such a detection scheme which can be treated as a temporal equivalent of diffraction-grating-based spectrometers.**

Optical coherence tomography (OCT) is a fast, noncontact, and noninvasive technique enabling high-resolution 3D imaging [1, 2]. In almost three decades since its inception, OCT has become widespread in various areas of biology and medicine with continual expansion in clinical applications for diagnostic and intraoperative purposes [3]. The quality of OCT images is practically limited by the ability of the detection unit to efficiently acquire light backscattered from an object. In a clinical setting, the photon budget becomes even smaller, because the intensity levels of light illuminating the imaged tissue must lie within ANSI standards or ICNIRP-guidelines. For example, for eyes, the safety level is of the order of 1.7 mW and 5 mW at wavelengths around 800nm and 1060 nm, respectively, as calculated for 10 s of continuous wave exposure [4]. In some situations, the permissible light power densities do not allow high-fidelity imaging, especially when the object under investigation is highly scattering or absorptive.

An approach of mapping a spectrum of an optical pulse to a temporal waveform by means of a long fibre spool and a single-pixel detector is called Dispersive Fourier Transformation [5]. It has already been used in OCT to time-stretch light from a supercontinuum source [6] and Ti:Sapphire laser [7] at the input of the interferometer and perform swept-source-like OCT at axial scan rates of up to 90 MHz.

Here, we propose a spectral detection scheme which is increasingly used in quantum optics to study spectral correlations of photons [8]. It is based on the principles of Dispersive Fourier Transformation - a fibre spool induces wavelength-dependent time delay and a single-pixel detector, which in this case is a single-photon detector, provides a fast and ultra-sensitive time acquisition. We show that such combination allows OCT imaging at light power levels which are at least five orders of magnitude lower than the safety standards. We report basic characteristics of such a detection system at 1550 nm and present images of a glass stack and an onion showing that the quality is comparable to the images obtained with standard OCT systems with similar axial resolution and imaging range characteristics.

An OCT system with the quantum-inspired spectral detection is presented in Fig. 1. Light with a central wavelength of 1550 nm and a total spectral bandwidth of 115 nm (MenloSystems T-Light) is inputted into a Linnik-Michelson interferometer through a fibre collimator FB1 (f=11 mm). The interference signal is then coupled into a 5-km long fibre spool (SMF28E, Fibrain) with a group velocity dispersion, $\beta_2$, equal to 23 $fs^2$/mm and then measured by a Superconducting Single-Photon Detector (SSPD) featuring a timing jitter of 35 ps. The SSPD is synchronized with the fast built-in photodiode placed in the light source. Because the fibre spool – through the phenomenon of dispersion – delays each wavelength by a different amount of time, time measurement performed by the SSPD effectively provides a spectrum of the light at the input of the fibre collimator FB2 (f=11 mm). Measured spectra are digitized by FPGA electronics (QuTag, QuTools gmbh) and saved onto a computer. Because the fibre spool's dispersion curve is not a linear function in wavenumber, a linearisation of the acquired spectra is performed [9].

To assess the performance of the detection system in the context of OCT imaging, a mirror was used as an object and its axial position was varied to introduce different optical path

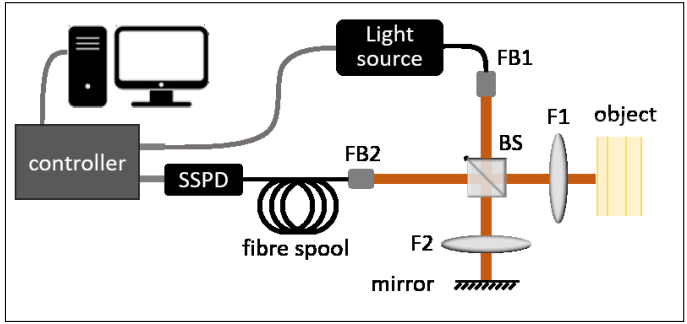

**Fig. 1.** Light propagates in a Michelson interferometer and is coupled into a fibre spool, which introduces wavelength-dependent time delay. Superconducting Single-Photon Detector (SSPD) performs time delay measurement for which a photodiode incorporated in the laser source unit provides a point of reference. FB1, FB2 – fibre collimators, F1, F2 – lenses (focal length of 50 mm), BS - beamsplitter

differences (OPD) in the interferometer. The axial resolution in air was around 17 µm and dropped from approximately 16.5 µm to 17.9 µm over the distance of 1.1 mm (Fig. 2d). As a reference, a spectrum of the light was measured with an optical spectrum analyzer (Fig. 2a) – the FWHM was 65 nm which at 1550 nm corresponds to a theoretical axial resolution of 16.3 µm. The 6-dB fall-off was calculated to be at 0.92 mm. The maximum imaging range, i.e. the maximum axial distance at which no aliasing is observed, was 5.1 mm. It means that although the detection system would allow detecting fringes at OPDs up to 5.1 mm, the sensitivity drops so fast that the fringes visibility becomes zero much earlier (compare Fig. 2b and 2c).

This behavior is analogous to the signal roll-off observed in traditional spectrometers [10]. The signal roll-off can be used to calculate the sensitivity fall-off, which is the decrease of the height of the peak in the A-scan with an increasing OPD. In traditional spectrometers, fall-off, $F_{gr}$, is approximated by the first three terms of Eq. 5 from [10] with the assumptions that the spectrum is distributed on the camera as a linear function of a wavenumber $k$ and that the spectrum itself is Gaussian:

$$F_{gr}(z) = \Delta x\, R\, e^{-\frac{a\, R^2 z^2}{4\ln 2}} \frac{\sin(\Delta x R z)}{\Delta x R z} \quad (1)$$

where $\Delta x$ is a pixel width of a camera sensor, $a$ - the FWHM of a single-wavelength spot size on the camera, $R$ - a reciprocal linear dispersion indicating the width of the spectrum that is the spectrum in wavenumber spread over 1 µm at the focal plane. The fourth term in eq. 5 from [10] is omitted here, because it is only responsible for generating Gaussian-shaped peaks in the model and does not contribute to the peaks' height decrease.

The first three terms of Eq. (1) representing a spatial case can be rewritten to a temporal form, $F_{temp}$, to describe a fall-off for fibre-based spectrometers:

$$F_{temp}(z) = \Delta t\, R_t\, e^{-\frac{\delta t\, R_t^2 z^2}{4\ln 2}} \frac{\sin(\Delta t R_t z)}{\Delta t R_t z} \quad (2)$$

$\Delta t$ is the Full Width at Half Maximum of the time jitter, and replaces the pixel size $\Delta x$ in Eq. (1). $R_t$ - the temporal analogue of a reciprocal linear dispersion - is the width of the spectrum in wavenumber spread over a time unit at the output of a fibre in fibre-based spectrometers. It is experimentally defined by $R_t = \Delta k/\Delta T$, where $\Delta k$ is the total width of the spectrum in wavenumber and $\Delta T$ is the pulse duration at the exit of the fibre. $R_t$ can be rewritten in terms of $\beta_2$ and fibre length, $L$, as $R_t = 1/(c\beta_2 L)$, where $c$ is the speed of light. $\delta T$ - whose spatial counterpart is the FWHM of a single-wavelength spot size on the camera - is a length of time corresponding to a spectral width equal to the spectral resolution of the fibre's DFT property. It can be calculated using eq. 21 in [11]: $\delta T = 2\sqrt{\pi \beta_2 L}$. $z$ is the OPD in the interferometer.

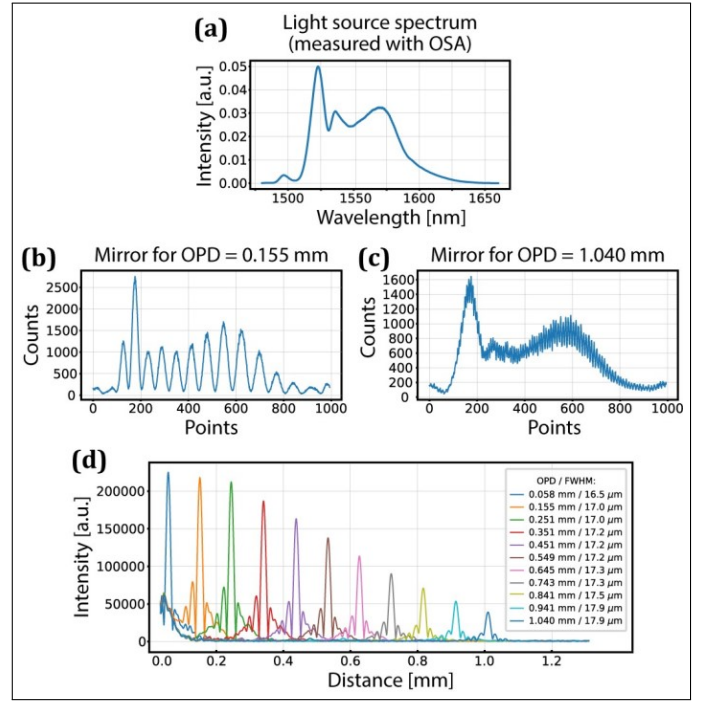

**Fig. 2. (a)** Spectrum measured by an optical spectrum analyzer, **(b,c)** Interference spectra at the optical path difference (OPD) of 0.155 mm and 1.04 mm show the decrease in fringe visibility similar to the one observed for traditional spectrometers [10], **(d)** Intensity of the spectra FFT for a mirror as an object for different OPDs of the interferometer. The 6-dB fall-off occurs at 0.92 mm and the axial resolution drops by 1.4 µm – from 16.5 µm to 17.9 µm – on a distance of 1.1 mm.

Eq. 2 describes a fall-off for linearly sampled spectra, which in DFT can only be achieved when two fibre spools are used [12], because the dispersion of the first fibre is compensated with the dispersion of the second fibre. To correctly simulate the fall-off for a single fibre spool, a more general Eq. 4 from [10] should be modified and used. In this formula, the spatial spectrum distribution, $x(k)$, is replaced with the temporal spectrum distribution $t(k)$.

The signal roll-off was measured for the lengths $L$=5 km and 3.5 km of a SMF28E fibre. The calculated fall-offs are presented in Fig. 3 as purple rectangles for the 3.5-km-long fibre and pink dots for the 5-km-long fibre. The experimental results are in fairly good agreement with the proposed model - solid and dashed theoretical curves in Fig. 3.

The spectral detection based on a fibre spool and a single-photon detector was used to image two kinds of objects: a stack of glasses (Fig. 4) and an onion (Fig. 5). The stack of glasses consists of a 50-µm thick quartz, a 460-µm thick sapphire and 500-µm thick BK7. Because the lateral size of the quartz was substantially smaller than the size of the sapphire, we decided to image the area on the edge of the quartz (Fig. 4). The left-hand side of images in Fig. 4 corresponds to just the sapphire and the right-hand side – to where the quartz lies on the sapphire. Fig. 4a presents a raw B-scan obtained after Fourier transforming linearised spectra, and Fig. 4b – the same B-scan processed to show the object without artefacts. On both images, one can discern the layer of quartz, an air gap between the quartz and the sapphire, the layer of sapphire and an air gap between the sapphire and the BK7. The second air gap was possible to see at the optical depth of around 1.4 mm despite the 0.92 mm 6-dB sensitivity fall-off. However, the increased sensitivity of the detection based on single-photon detectors allowed the appearance of "parasitic" peaks, which are associated with the interference of photons reflected from every surface of the object. Because for a layered structure these peaks are always placed at the same respective distances from the 0 OPD point, they can be removed (as depicted in Fig. 4(b)) by subtracting a mean spectrum from every spectrum in the B-scan file. We should note here that these "parasitic" peaks are always present in the B-scan, irrespective of the angle at which the glasses are positioned with respect to the beam.

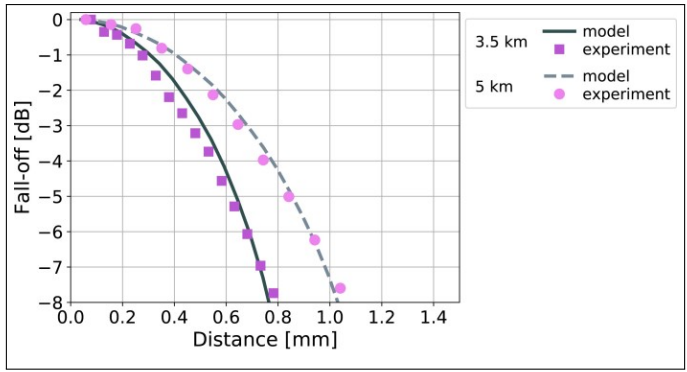

**Fig. 3.** The theoretical and experimental fall-offs calculated for two lengths of a SMF28E fibre: 3.5 km and 5 km. The STD of the jitter, $\Delta t$, in the calculations is 35 ps. The temporal equivalent of spectral resolution of the fibre, $\delta T$, was calculated to be 32 ps for the 3.5-km-long fibre and 38 ps for the 5-km-long fibre. The fall-off, F, was calculated as 10log(F), where F is normalised.

The same removal algorithm cannot be used on objects whose structure is more complicated than parallel layers such as biological specimen. A piece of onion (Fig. 5) was imaged to partially visualize the problem. Fig. 5 depicts a mean of 10 images at one cross-section of the object. The self-interference terms can be seen close to the 0 OPD point. One can discern a cellular structure of the object, but the image is very grainy, even after averaging. Also, the quality of the image can be degraded by the fact that the structure of the object overlaps the area with the self-interference terms. This problem might be remedied if the object is placed at a larger OPD. Nevertheless, the image quality is similar to the quality of images obtained by standard Spectral OCT systems with comparable imaging parameters, but orders of magnitude higher input powers [13].

We presented a detection scheme which is widely used for spectral measurements in quantum optics and implemented it for OCT imaging. This detection scheme allows imaging with extremely reduced light intensity levels, around 10 pW, and at the same time, with the same quality as compared to standard OCT systems. However, the blessing of a very sensitive detection enabled by single-photon detectors proves to be a curse due to the omnipresent self-interference peaks. Whereas it is very easy to get rid of these peaks for well-defined structures such as glass, these parasitic peaks become problematic for more complicated objects such as biological specimen. In such a case, an efficient artefact removal can possibly be achieved with an algorithm based on deep learning which has recently proven its superiority in a vast number of problems associated with imaging [14–17].

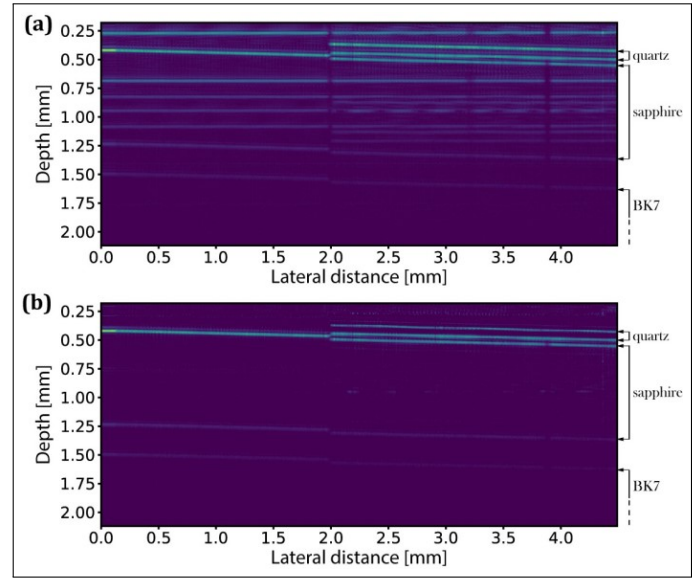

**Fig. 4.** An image of a stack of glass: quartz (only on the right), air gap, sapphire and an air gap between sapphire and BK7. **(a)** A B-scan showing an increased sensitivity in detecting interference between photons reflected from every surface of the object, **(b)** The same B-scan where the additional peaks were removed numerically. Layers way below the 6-dB fall-off distance are still visible.

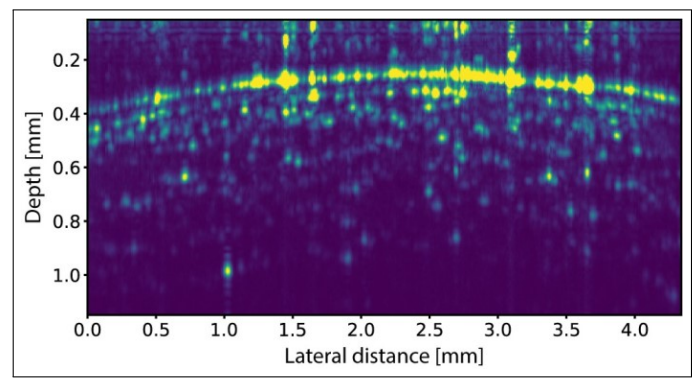

**Fig. 5.** An image of an onion slice. An average of 10 B-scans acquired at the intensity level of approx. 10 pW.

**Funding Information.** SMK and FV acknowledge New Zealand Royal Society Marsden fund (UoA1509); The Dodd-Walls Centre for Photonic and Quantum Technologies (New Ideas Fund). PK acknowledges financial support by the Foundation for Polish Science (FNP) (project First Team co-financed by the European Union under the European Regional Development Fund).